\documentstyle[11pt,aaspp4]{article}

\def\gtrapprox{\;\lower 0.5ex\hbox{$\buildrel >
    \over \sim\ $}}             
\def\lessapprox{\;\lower 0.5ex\hbox{$\buildrel < \over \sim\ $}}

\def\kms{~km s$^{-1}$}
\def\mic{~$\mu$m\/}

\begin{document}
\title{New CO and Millimeter Continuum Observations of the $z = 2.394$ 
Radio Galaxy 53W002}

\author{Danielle Alloin}
\affil{ESO, Casilla 19001, Santiago, Chile}
\centerline{Email: dalloin@eso.org}

\author{Richard Barvainis}
\affil{National Science Foundation, 4201 Wilson Boulevard, 
Arlington VA 22230}
\centerline{Email: rbarvai@nsf.gov}

\author{Stephane Guilloteau}
\affil{IRAM, 300 Rue de la Piscine, 38406 Saint Marten d'H\`eres, France}
\centerline{Email: guillote@iram.fr}

\begin{abstract}
The $z=2.39$ radio galaxy 53W002 lies in a cluster of Ly$\alpha$
emission line objects and may itself be undergoing a major burst of
star formation.  CO(3--2) emission, at 102 GHz, was detected from
53W002 by Scoville et al.\ (1997a), who also reported a possible
30 kpc extension and velocity gradient suggesting a rotating
gaseous disk.  In this paper we present new interferometric
CO(3--2) observations which confirm the previous line detection
with improved signal-to-noise ratio, but show no evidence for source
extension or velocity gradient. 
The compact nature of the
CO source and the molecular mass found in this object are similar
to luminous infrared galaxies and other AGNs previously studied.

\end{abstract}
\section{Introduction}

  The distant, compact, narrow-line radio galaxy 53W002, at $z = 2.39$, has 
been the subject of some interest recently because it may be a 
relatively young galaxy undergoing a major episode of star formation 
(Windhorst et al 1991), and because it appears to be the dominant galaxy 
in a field of 17 faint Ly$\alpha$ emission line objects, of which 7 have
been confirmed to be at a redshift close to that of 53W002. This is the 
most distant such cluster known ( Windhorst et al 1991, Pascarelle et al 1996a, 
1996b), and one motivation for its further study is to test the formation of 
giant galaxies through the assemblage of sub-galactic star-forming clumps 
(Windhorst et al 1998, Pascarelle et al 1998). In this respect, analysis of 
the molecular content and distribution in the dominant galaxy of this cluster, 
53W002, might bring some new input into the discussion. 
   
   To this end, Yamada et al.\ (1995) reported a tentative detection
of CO(1--0) emission from 53W002 using the Nobeyama 45m telescope,
which was followed by an interferometric measurement of CO(3--2)
by Scoville et al.\ (1997a; hereafter S97a) using the Owens
Valley Millimeter Array.   Of particular interest in the latter
study was that the CO source appeared to be extended by $\sim 30$
kpc, although the signal-to-noise ratio of the measurement was not
very high.  In other known luminous CO sources, such as Arp 220 and
the Cloverleaf quasar (H1413+117), the molecular gas appears to
be concentrated in the inner few hundred parsecs of the galactic
nucleus rather than being distributed on kpc scales (Scoville et
al.\ 1997b; Kneib et al.\ 1998; Downes \& Solomon 1998).  As of
this writing there are 9 well-documented detections of CO at high
redshift (recently summarized by Combes et al.\ 1999), including
53W002, which is one of only two in this list that appears not to
be gravitationally lensed.

The presence of abundant molecular gas can be an indicator of
widespread star formation or nuclear starbursts, and can also be a 
signpost for the fueling of an active nucleus following a merger
or interaction between galaxies.   Since 53W002 has both an active
nucleus and apparent extended star formation, as revealed by its 
extension in the  Ly$\alpha$ emission line, and therefore may be under 
construction via mergers,  observations of molecular lines  
are  clearly germane to understanding the system.  
Given the relevance, and potential uncertainties, of the previous  
CO measurements, we undertook new observations of 53W002 in the CO(3--2)
line at 3mm and in the continuuum at 3mm and 1.3mm, using the IRAM 
Plateau de Bure Interferometer, which we report here.

\section{Observations and Results}

The observations were performed with the Plateau de Bure Interferometer of 
the Institut de Radio Astronomie Millimetrique over a number of sessions from 
September 1996 to November 1997.  We searched for the CO(3--2) line emission
using two central frequency setups, 101.945 and 102.195 GHz, for a total of
55 hours. During part of this time (~20 hours) data at 1.3mm were recorded 
simultaneously to probe the object's continuum at this frequency. The total 
frequency bandwidth was 512 MHz, 50 channels, and typical system 
temperatures were 130 K at 3mm and 500 K at 1.3mm. A number of sources were
used to track variations in the phase and gain. The source 1739+522 
was used for flux calibration which could be achieved within an accuracy of 
5\%. The various configurations of the interferometer array, including 4 or 5 
15m antennas, covered baselines from short (24m) to long (408m). 
Combining the whole data set results in a final  
synthesized beamsize of $5''\times 6''$ (natural weighting) at
PA=90$\arcdeg$, and covers a field-of-view of $50''$ . 
For the results presented hereafter, the velocity origin is at frequency 
101.8845 GHz. Because of the two different central frequency settings, the 
(u-v) plane coverage is not identical at all frequencies across the CO(3--2) 
line-profile, which results into a slightly changing spatial resolution 
across the line-profile: for channels centered at -823, -588 and -353\kms , 
the spatial resolution is better than for channels centered at -117, +117 
and +353\kms . 

  CO(3--2) emission was detected with an integrated flux of
$1.20\pm 0.15$ Jy\kms . The line profile is shown in Figure
1 (velocity resolution 118\kms ), channel maps in Figure 2
(velocity resolution 235\kms ), and the velocity-integrated 
CO map in Figure 3 (700\kms\ window).  The maps in Figure 2
have an RMS uncertainty of 0.45 mJy beam$^{-1}$, and for Figure 3 
the map RMS is 0.25 mJy beam$^{-1}$.  The best 
fit position for the integrated CO
source  is RA(B1950) = $17^h\ 12^m\ 59^s.72$ and 
Dec(B1950) = $50\arcdeg\ 18'\ 51.9''$, with an uncertainty of $0.5''$
in each coordinate.   This uncertainty is from two components added
in quadrature,  one due to phase and baseline errors amounting to $0.3''$,
and a second from the map noise of $0.4''$.  For the 235\kms\ 
channel maps of Figure 2, the positional uncertainty in the channels
at $-353$\kms\ and $-118$\kms\ is $0.8''$, and that of the channel
at $+117$\kms\ is $1.1''$.           
 
We find no significant CO emission from any of the Ly$\alpha$ galaxies
in the field surrounding 53W002.

At 3mm (101 GHz) the measured continuum flux density, using line-free
channels, is $0.1\pm 0.2$ mJy, yielding a $3\sigma$ upper
limit of $< 0.6$ mJy.  At 1.3mm (230 GHz), using both sidebands,
the measured flux density is $1.8\pm 0.9$ mJy.  In collaboration with
R.\ Antonucci and T.\ Hurt, we also obtained
a 1.3mm measurement using the IRAM 30m telescope 19-channel
bolometer of $1.6\pm 0.5$ mJy (February 1997).  Combining this
with the PdBI result, the weighted average is $S_{\rm 1.3mm} = 1.7\pm 0.4$
mJy.  Any contribution from the centimeter wavelength synchrotron
spectrum should be negligible at millimeter wavelengths (see Figure 4 of S97a).

Measured CO and continuum parameters and derived luminosities are summarized 
in Table 1.

\section{Discussion}

\subsection{CO Emission}

First we compare our results with those of S97a.  
The integrated CO fluxes are in good agreement
between the two experiments :  $1.20\pm 0.15$ Jy\kms\  found
here versus $1.51\pm 0.20$ Jy \kms\ found by S97a.
Yet, the error quoted by S97a seems to be underestimated for
the following reason: the RMS given in their Figure 3 is 
0.44 mJy beam$^{-1}$, while the CO line flux is integrated
over a $5''$ aperture which comprises two independent beams.  Hence
the error bar on the line flux should read $\pm 0.3$
Jy\kms\ ($0.44~ {\rm mJy} \times \sqrt 2  \times 540$\kms ) 
rather than $\pm 0.2$ Jy\kms\ as quoted, yielding a $5\sigma$
detection only.  The derived line redshifts between this experiment and 
S97a are consistent within the errors.

S97a noted a positional shift with velocity in their channel
maps, leading to the conclusion that the source was extended
and possibly rotating.  From the velocity-integrated CO map they
derived a deconvolved CO(3--2) source size of $5.7''\times 1.7''$
(FWHM for an elliptical gaussian), or about $30\times 9$ kpc ($H_0 =
75$\kms kpc$^{-1}, q_0 = 0.5$), at position angle $120^{\rm o}$.
We can neither confirm nor exclude a source of these dimensions,
based on our integrated CO map alone.  The source can equally
well be unresolved, $2''$ in diameter, or $5.7''\times 1.7''$
at any position angle.  S97a note that ``In view of the limited
signal-to-noise ratio in the maps, these size estimates are highly
uncertain; however, it is clear that the emission is resolved,
since the centroid is shifted in the channel maps.''

This shift in the centroid is the crucial evidence for extension,
but our maps do not show it.  For direct comparison we plot the same
channel maps in our Figure 2 as those of S97a (also their Figure
2): $-353$, $-118$, $+118$, and $+353$\kms\  (plus we include
maps at $-588$\kms\ and $-823$\kms ).  Between the $-118$\kms\
and $+118$\kms\ maps of S97a there is a position shift of the
peak emission of roughly $5''$, whereas in our channel maps (the
three with significant emission) there is no shift in the spatial
peaks or centroids to within our relative astrometric accuracy
of $\approx 1''$.  Note that emission is also seen at $-353$\kms\
in our data, but not in the maps of S97a.   The lower contours of
our integrated CO map (Figure 3) are distorted somewhat toward the
southwest, but this distortion is not significant within the map
noise uncertainties.  We conclude from our data that a velocity
gradient with position such as that seen by S97a is not supported,
thereby casting serious doubt on the reality of any extension to the
CO emission in 53W002, at least on the scale of a few arcseconds.
The evidence from the Plateau de Bure is consistent with a point
source spatially coincident at all velocities.

With regard to the Yamada et al.\ (1995) CO(1--0) observation, S97a
noted that the CO(1--0) flux is far too strong to be a counterpart of
the observed CO(3--2), given the normal range of CO excitation seen
in galaxies.   The same conclusion applies to our own measurement,
which is even somewhat weaker.  The CO(1--0) as measured by Yamada
et al.\ is an order of magnitude or more too strong to be consistent
with the CO(3--2) emission.

The mass of molecular hydrogen in 53W002, as derived from the CO
luminosity, depends on one's choice of the $L'_{\rm CO} \rightarrow
M({\rm H}_2)$ conversion factor, $\alpha$.  The standard value
for Galactic molecular clouds based on CO(1--0) data, $\alpha =
4 M_{\sun}$ (K \kms\ pc$^{-2}$)$^{-1}$, is almost certainly too
high for dense starbursts and active galactic nuclei.  Downes \&
Solomon (1998) find a value 5 times lower in a study of extreme
starbursts in ultraluminous infrared galaxies, and Barvainis et
al.\ (1997) derive a value 10 times lower in a multi-line study
of the Cloverleaf quasar.  For $L'_{\rm CO} = 1.5\times 10^{10}$
K\kms\ pc$^2$ and using the Downes \& Solomon conversion factor,
the molecular hydrogen mass would be roughly $1.2\times 10^{10}
M_{\sun}$, while using the Barvainis et al.\ factor it would be
around $6\times 10^9 M_{\sun}$.
The CO luminosities of 53W002 and the Cloverleaf are
quite comparable, once account is taken for lensing in the latter.
If the CO magnification factor is 10 for the Cloverleaf, then its
intrinsic CO luminosity is $L'_{\rm CO} = 1.0\times 10^{10}$ K\kms\
pc$^2$ (Barvainis et al.\ 1997), compared with $1.5\times 10^{10}$
K\kms\ pc$^2$ for  53W002 (Table 1).

\subsection{Astrometric considerations}

In Table 2 we provide a compilation of available astrometric measures
of 53W002 at 8.4 GHz (Windhorst et al.\ 1991), 15 GHz (S97a),
optical continuum (Windhorst et al.\ 1991), and CO line emission
(S97a and current results).  The positions summarized in Table 2
show that the location of the CO source, which is identical in the
three velocity channel maps of the IRAM data set, is consistent
with that derived from the {\it integrated} map in S97a.  The CO
source location also coincides, within the error-bars, with the VLA
8.4 GHz source position (Windhorst et al.\ 1991), while in right
ascension it is more than $2 ''$ off the position reported for the
VLA 15 GHz source (S97a). This offset is well beyond the quoted
error-bars for the 8.4 and 15 GHz VLA source positions.   For the
current discussion we shall retain  the 8.4 GHz positional measure
which appears to be more precise.  A similar offset around $1.7''$
in right ascension is also found between the CO source location and
the optical (rest ultraviolet) continuum source (Windhorst et al.\
1991). This offset however may simply reflect the slight mismatch
known to exist between radio and optical reference frames and is not
considered to be as significant. Indeed, recently Windhorst et al.\
(1998) have noted for 3 other AGNs in the field around 53W002 a
mismatch of the same order of magnitude between their radio (VLA 8.4
GHz) and optical positions: this reconciles the 8.4 GHz and optical
continuum positions for 53W002, but, in our view does not explain
the disagreement between the VLA 8.4 GHz and 15 GHz source locations.

  From this astrometric analysis, we conclude that the CO and 8.4 GHz sources 
are coincident and consistent with the optical continuum source location,
according to the above mentioned shift between radio and optical reference 
frames. This leads us to conclude that the CO source is related to the
AGN located at the heart of 53W002 (Windhorst et al.\ 1991).

\subsection{Relation of the CO line emission to other components}

  As 53W002 has been considered a key object for testing scenarios of 
galaxy formation, it is important to establish its properties solidly.

An important question for interpreting the {\it evolutionary stage}
of 53W002 is to determine whether the CO line emission is mostly
related to the AGN and surrounding activity or to intense star
formation across the body of the galaxy. Given the fact that the
higher signal-to-noise ratio IRAM data set in CO does not support the
existence of a velocity gradient across the CO line emitting region,
hence an extension, the validity of associating part of the CO line
emission with infalling sub-galactic clumps still in the body
of a forming galaxy (Windhorst et al.\ 1998) should be questioned.

On the other hand, taking into account the compact configuration of
molecular gas systems seen in most ultraluminous infrared galaxies
and AGNs, an association of the CO line emission with the various
components identified by Windhorst et al.\ (1998) within the central
$1''$ HST image seems more plausible. These include the AGN per-se
(unresolved source with size $\lesssim 0.06''$), a compact core
with $r_e \sim  0.05''$, a more extended envelope with $r_e \sim
0.25''$ and two ``blue clouds'' roughly colinear across the nucleus,
along PA$\sim 90\arcdeg$, at distance $0.45''$ W and $\sim 0.2''$
E, which are well contained within the 8.4 GHz radio source. The
latter also shows an elongation along PA$= 90\arcdeg \pm 3\arcdeg$
on a scale of $\sim 0.5''$.

  Obviously none of the components identified in the HST data can be
spatially disentangled relative to the CO line contribution.
As noted above however, the CO luminosities of 53W002 and the
Cloverleaf quasar are comparable.  In the Cloverleaf the molecular
gas was found to reside in the inner 150 pc region by Kneib et al.\
(1998), who argued that this is the signature of the molecular/dusty
torus surrounding the active nucleus. In their early imaging and
spectroscopic analysis, Windhorst et al.\ (1991) concluded that the
AGN in 53W002 exhibits a continuum luminosity close to the boundary
with QSOs.  By analogy with the Cloverleaf (and ultraluminous
infrared galaxies in general) it can be argued that the CO line
emission detected in 53W002 comes mostly from the molecular/dusty
torus of the AGN itself. Interferometric data at higher resolution
are required to resolve this point.

\subsection{Continuum  Emission}

   Hughes \& Dunlop (1998) reported a
$3\sigma $ upper limit of 3.3 mJy for 53W002 in the continuum at
800\mic .  Combining this upper limit (corrected to 800\mic\ assuming
a submillimeter spectral index of 3.5) and the value $S({\rm CO})
= 1.2$ Jy\kms\ from Table 1, we find that $S({\rm CO})/S_{800}
> 294$\kms .  For four other high-$z$ sources with both CO(3--2)
and 800\mic\ continuum detections, this ratio ranges from 88\kms\
to 165\kms .  Therefore, in terms of CO to dust ratio, the case of 53W002
looks somewhat peculiar.

  Our measurement of $1.7\pm 0.4$ mJy at 1.3mm
is well above the extrapolation of the centimeter wavelength
spectrum (S97a), suggesting the presence of an extra component.
Because upper limits only are available for the continuum at 3mm and
850\mic , the nature of this extra component cannot be constrained
further.  If it is dust at about 50K, the upper limit to the mass of
dust inferred from the measure at 850\mic\ is $M_{\rm D} < 7\times
10^7 M_{\sun}$ for the cosmological parameters chosen in this paper
(Hughes and Dunlop 1998).

\section{Conclusions}

This new set of observations of CO(3--2) line emission in the
radio galaxy 53W002, collected with the IRAM Plateau de Bure
interferometer, is of an improved (8 sigma) signal-to-noise ratio
relative to previous studies.  The individual channel maps (averaged
over 235\kms ) do not show any positional velocity shift and hence
do not support an extension of the CO line emitting region beyond
$3''$ (FWHM).  Within the quoted uncertainties, the CO(3--2) source
is found to be positionally coincident with the VLA 8.4 GHz source,
as well as with the HST optical continuum (rest ultraviolet) image of
53W002. We conclude that a large fraction of the CO line emission is
likely related to the AGN itself or its immediate surroundings. The
CO line profile has a FWHM = $420\pm 40$\kms , while the total
CO line flux is found to be $1.20 \pm 0.15$ Jy\kms . Assuming a
figure $\alpha = 0.4 - 0.8 M_{\sun}$ (K\kms\ pc$^{-2}$)$^{-1}$
for the conversion factor from $L'_{\rm CO}$ to M(H$_2$), which
better applies to dense regions around the centers of galaxies,
we derive a molecular mass in the range $6\times 10^9 - 1.2\times
10^{10} M_{\sun}$. This value is comparable to the one found within
the 150 pc region in the Cloverleaf quasar.  Therefore, the CO line
emission observed in 53W002 could be entirely related to the AGN
of this galaxy. New measurements of the millimeter continuum of
53W002 at 1.3mm, as well as upper limits for the continuum at 3mm
and 850\mic , imply the presence of an extra component in addition
to the synchrotron source responsible for the centimeter spectrum.
If interpreted in terms of dust, then the implied gas to dust ratio
in 53W002 is larger than $10^2$.

\acknowledgments 
We are grateful to the referee, David Hughes, for very helpful
comments, and to the IRAM staff for excellent support during the
observations performed in service mode.

\newpage
\vglue 1.0truecm
\centerline {\bf FIGURE CAPTIONS}
\figcaption{
CO(3--2) line profile, smoothed to 118\kms\ resolution.  Upper scale:
velocity relative to redshift of $z = 2.394$ (central frequency 101.8845
GHz).  Lower scale: observed frequency.
\label{fig1}}
\figcaption{
CO(3--2) channel maps averaged 235\kms\ velocity ranges.  The central
cross marks the position of the VLA 8.4 GHz radio source (see Table 2).   
The contours are $-3$, $-2$,
+2, +3, and +4
times 0.45 mJy beam$^{-1}$ ($1\sigma$).  The velocities are relative
to $z = 2.394$, and the $6''\times 5''$ naturally weighted beam
at PA$= 90\arcdeg$ is shown.  The coordinates are J2000.0.       
\label{fig2}}
\figcaption{
Velocity-integrated CO(3--2) map of 53W002.  Contours range from
$-3$, $-2$, and +2 to +6 times the $1\sigma$ RMS of 0.25 mJy beam$^{-1}$.
The CO is consistent with a point source ($< 3''$ FWHM); the 
southwest extension of the lower contours is not statistically 
significant.  No emission from the Ly$\alpha$ emission line 
objects in the optical field is apparent.  The position of the 
central cross is as in Figure 2.  The coordinates are J2000.0
\label{fig3}}
\newpage
\begin{deluxetable}{lcc}
\tablenum{1}
\tablewidth{0pt}
\footnotesize
\tablecaption{CO(3--2) Line and Continuum Parameters}
\tablehead{
\colhead{Parameter} &\colhead{Value} 
&\colhead{Units}
}
\startdata
$z_{\rm CO}$  &$2.3927\pm0.0003$ & \\
$\Delta v_{\rm FWHM}$ & $420\pm 40$ & \kms \\
$S({\rm CO})$ &$1.20\pm 0.15$ & Jy\kms \\
$L'_{\rm CO}\tablenotemark{a}$ &$1.48\times 10^{10}$ & K\kms pc$^2$\\
$S$(102 GHz)& $0.1\pm 0.2$ & mJy\\
$S$(230 GHz)& $1.7\pm 0.4$ & mJy\\
\tablenotetext{a} {Assumes $H_0 = 75$ km s$^{-1}$ Mpc$^{-1}$, $q_0 = 0.5$} 
\enddata 
\end{deluxetable} 

\begin{deluxetable}{lcllc}
\tablenum{2}
\tablewidth{0pt}
\footnotesize
\tablecaption{Positions for various components in 53W002}
\tablehead{
\colhead{Component} &\colhead{Waveband} 
&\colhead{RA(B1950)}
&\colhead{Dec(B1950)}
&\colhead{Reference\tablenotemark{a}}
}
\startdata
CO (IRAM)& 3mm 
& $17^{\rm h} 12^{\rm m} 59^{\rm s}.72\pm 0.05$ &
$50^{\rm o} 18' 51''.9 \pm 0.5$ & 1\\
CO (OVRO)\tablenotemark{b}& 3mm
& $17^{\rm h} 12^{\rm m} 59^{\rm s}.68$ &
$50^{\rm o} 18' 51''.7 $ & 2\\
Radio Continuum& 8.4 GHz
& $17^{\rm h} 12^{\rm m} 59^{\rm s}.76\pm 0.01$ &
$50^{\rm o} 18' 51''.7 \pm 0.1$ & 3\\
Radio Continuum& 14.9 GHz
& $17^{\rm h} 12^{\rm m} 59^{\rm s}.86\pm 0.01$ &
$50^{\rm o} 18' 51''.3 \pm 0.1$ & 2\\
Optical Continuum& Gunn g \& r
& $17^{\rm h} 12^{\rm m} 59^{\rm s}.83\pm 0.04$ &
$50^{\rm o} 18' 51''.3 \pm 0.4$ & 3\\
\tablenotetext{a} {References: (1) This paper; (2) S97a (3) Windhorst
et al.\ (1991)}
\tablenotetext{b} {Position estimated from the center of the 
peak contour of the integrated CO emission 
from Figure 3 of Scoville et al.\ (1997a).  Positional uncertainties
unavailable.}
\enddata
\end{deluxetable} 

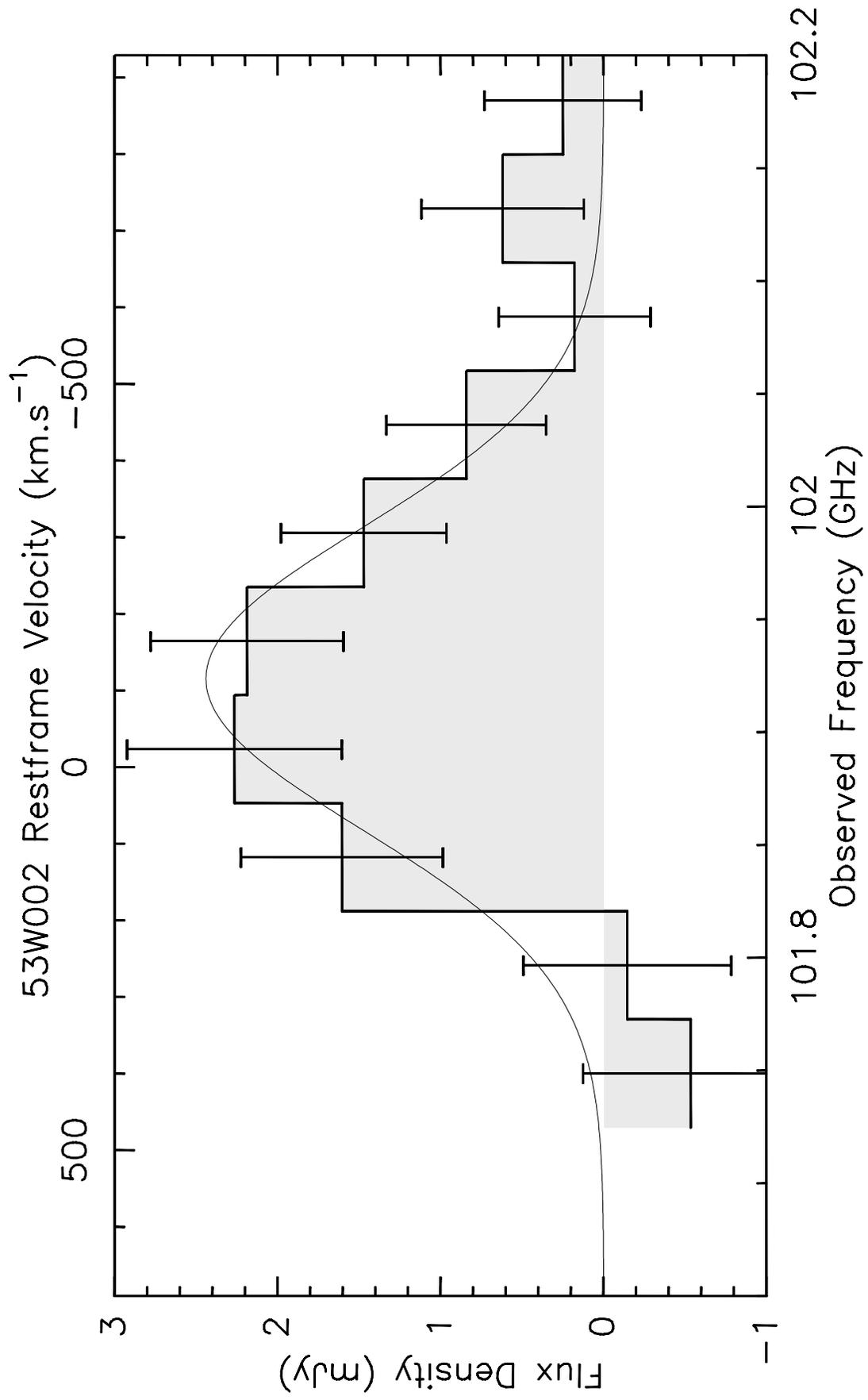
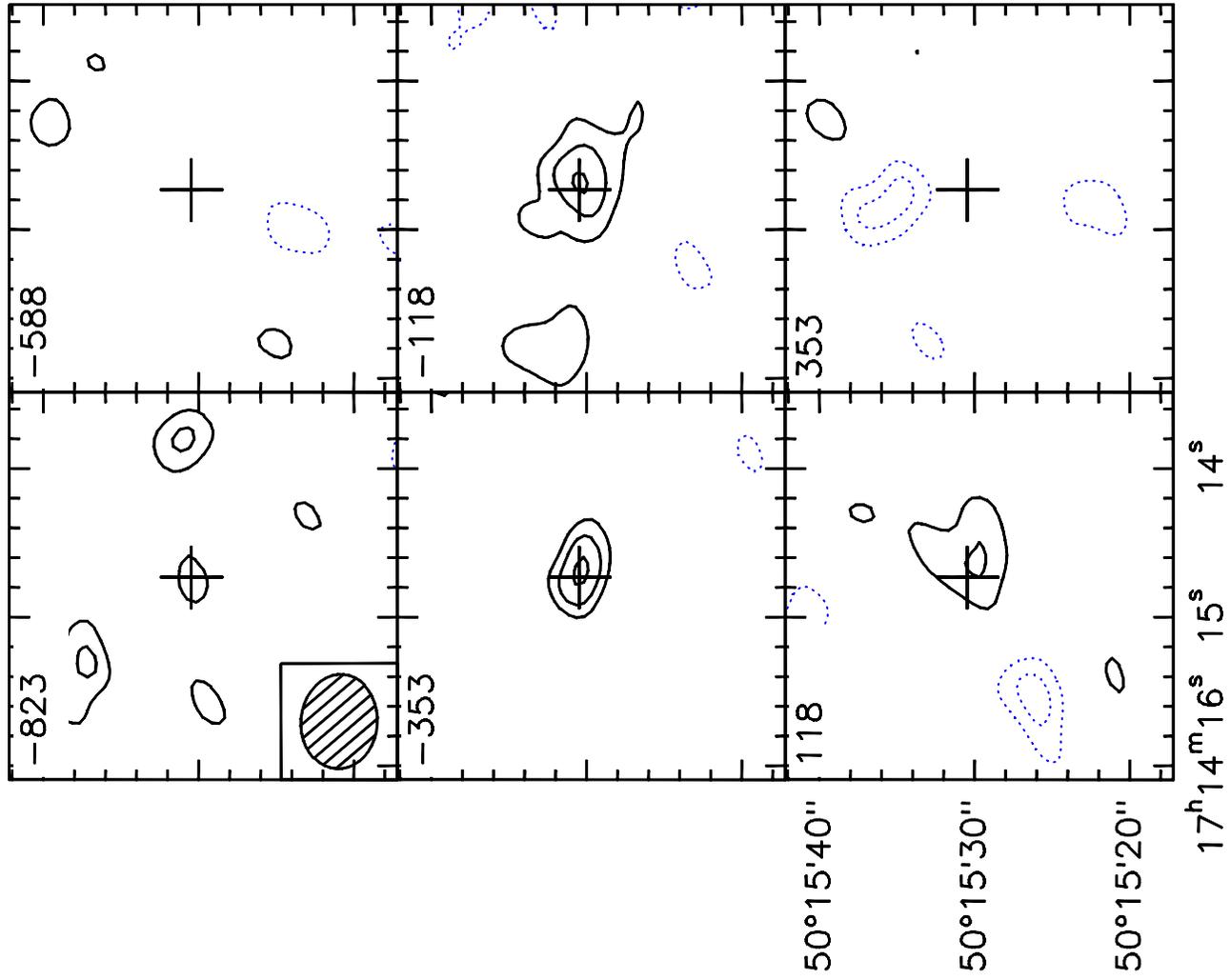
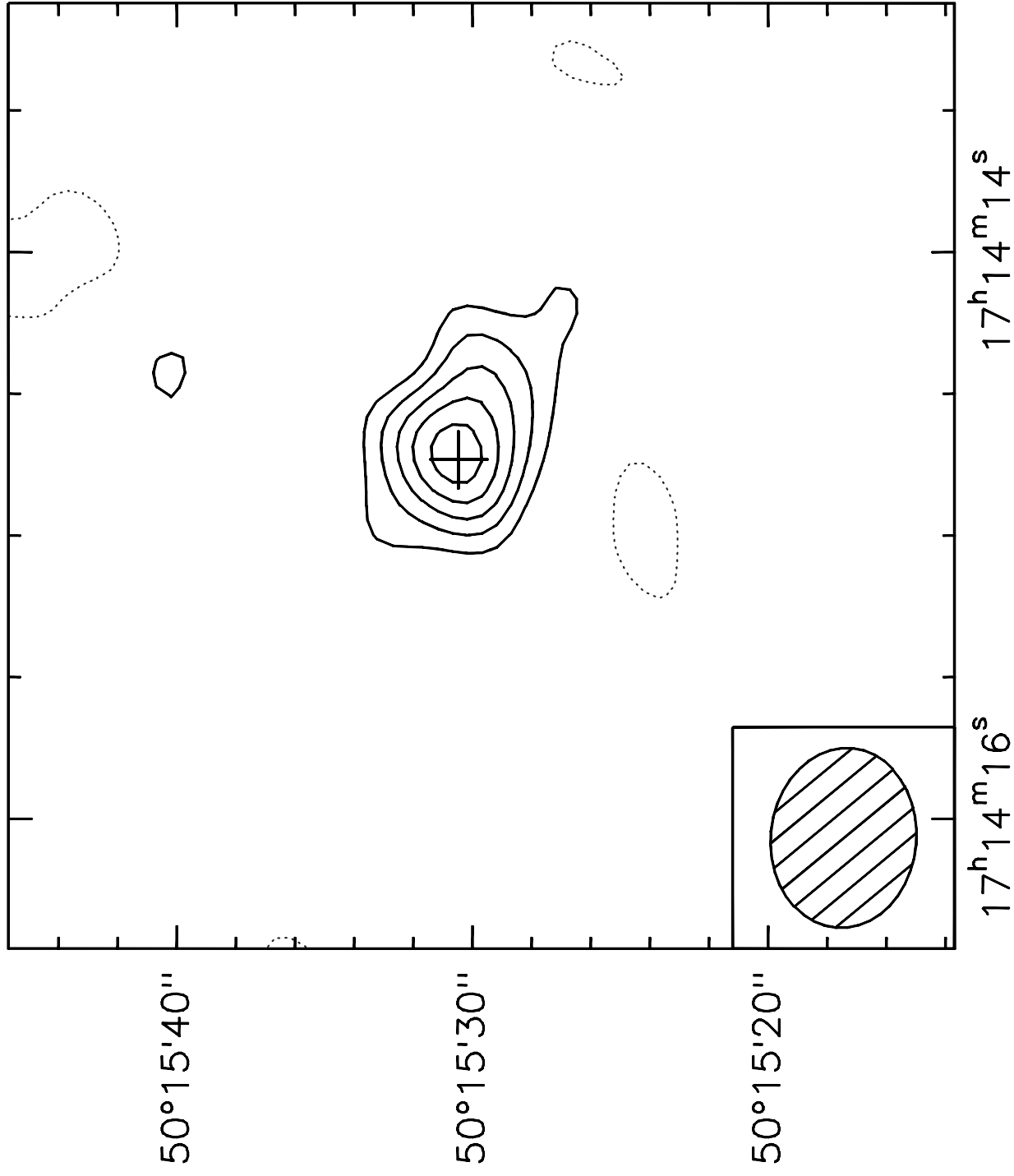
\end{document}